# Exciton Mott Transition in Two-Dimensional Semiconductors

Yiling Yu[1], Guoqing Li[1], Linyou Cao[1,2,3*]

[1]Department of Materials Science and Engineering, North Carolina State University, Raleigh, North Carolina, 27695, United States; [2]Department of Physics, North Carolina State University, Raleigh, North Carolina, 27695, United States; [3]Department of Electrical and Computer Engineering, North Carolina State University, Raleigh, North Carolina, 27695, United States

**Abstract**

Exciton many-body interaction bear great implication for application in advanced photonic devices and quantum science and technology such as quantum computing, but the fundamental understanding about exciton many-body interaction is very limited. Here we provide numerous new insights into the fundamentals of exciton Mott transition (EMT), a manifestation of exciton many-body interaction evidenced by the ionization of excitons into a plasma of unbound electrons and holes, *i.e.* electron-hole plasma (EHP), by taking advantage of the unique properties of two-dimensional (2D) semiconductors like monolayer $MoS_2$. We clarify long-standing controversies on the continuousness and criteria of EMT, quantify the charge carrier distribution among the co-existing exciton and EHP phases, establish correlation between the emission features and charge densities of EHP, and elucidate the physical state of EHP charge carriers as nanoscale electron-hole complex rather than individually free charges. These results lay down a foundation for furthering the studies of exciton many-body interaction and also for utilizing the interaction in quantum science/technology and the development of advanced optoelectronic devices.

* Correspondence should be addressed to: linyoucao@gmail.com

The many body interaction of quantum particles or quasi-particles such as electrons, phonons, excitons, and spins is one of the most exciting topical research areas of modern physical science and engineering, as it underlies many major scientific discoveries and technological breakthroughs, such as superconductivity[1], Bose-Einstein condensation[2], topological insulator[3], quantum entanglement[4], and quantum computing[5]. The many-body interaction of excitons, which consist of pair-wise electrons and holes bound by Coulomb attraction, is particularly intriguing.[6,7] Unlike other basic particles or quasi-particles such as electrons or phonons, which are inherent to matters, excitons can only exist with the presence of external excitation, either optical or electrical. This dependence on external excitation offers opportunities to externally control quantum many-body interactions that are difficult with other particles.

Exciton many-body interaction may be best manifested by the phase transition of excitons.[8-13] Intuitively, excitons at a low density behave like free gas molecules due to little exciton-exciton interaction and hence are also referred as free excitons (FEs). With the density increasing, the exciton many-body interaction becomes prominent to screen the pair-wise Coulomb attraction, which may eventually enable exciton Mott transition (EMT) with excitons ionized into a plasma of unbound electrons and holes, *i.e.* electron-hole plasma (EHP).[14-18] At the highest density, even stronger exciton many-body interaction can cause EHP or excitons to undergo a gas-liquid phase transition into a liquid-like state, *i.e.* electron-hole liquid (EHL).[9,11,16,19-21] Despite the well-recognized intuitive picture, quantitatively understanding for the evolution of exciton phases with increase of exciton density, in particular for EMT, has remained limited. Long-standing controversy or ambiguity remains for even the most fundamental issues of EMT. For instance, is the EMT an abrupt or continuous process? What is the criteria for the onset of EMT? How do the photo-generated charge carriers distribute among the co-existing exciton and EHP phases? What are the spectroscopic features of EHP and how the emission feature could be quantitatively correlated to the charge density? How is the EHP different from the unbound charges resulting from the regular entropy-driven exciton ionization? The key challenge for quantitatively understanding EMT lies in the difficulty to deterministically distinguish the signals from co-existing excitons and EHP.[14,16,22] The emission of these species in conventional semiconductors are very close in energy with difference usually less than 5 meV and thus difficult to separate.[16,22] Additionally, the external excitation (such as laser) required to enable EMT often induces inhomogeneity inside semiconductors due to the materials' non-trivial dimension, which can broaden the spectral distribution of emission from excitonic species and makes the differentiation more difficult.

Two-dimensional (2D) semiconductors provide an ideal platform for the fundamental studies of EMT due to unique properties.[23] The extraordinary binding energy of 2D semiconductor can enable larger energy separation between excitons and EHP, and the atomically thin dimension minimizes laser-induced inhomogeneity. Indeed, EMT have been recently reported at 2D semiconductors.[15,17,18] In this work, we focus on acquiring new insight into the fundamentals of EMT by taking advantages of 2D semiconductors. More specifically, we clarify the long-standing controversy and ambiguity on many critical issues, such as the continuousness and criteria of EMT, the distribution of charge carriers among co-existing exciton and EHP phases, the physical state of charge carriers in EHP, and the correlation between PL spectral features and charge density of EHP.

Without losing generality, we focus on $MoS_2$, more specifically, suspended single crystalline monolayer $MoS_2$ flakes. The suspension is to eliminate substrate effects for the simplicity of data interpretation as the presence of substrates can complicate the dynamics of excitons.[24,25] The single crystalline flakes were synthesized using well-established chemical vapor deposition (CVD) process,[26,27] and then transferred onto quartz substrates with pre-patterned holes in size of several micrometers as illustrated in Fig. 1b inset. We confirmed the composition and high crystalline quality of the flakes through the preparation process as how we did in previously studies.[11,28] Photoluminescence presents a powerful tool for the studies of exciton phase transition. We previously demonstrated significant changes in the PL of suspended monolayer $MoS_2$ upon the phase transition of excitons.[11] Fig. 1 shows the PL's spatial and spectral features collected from suspended $MoS_2$ monolayer with different incident power densities. The result is similar to what we reported previously.[11] Both the PL's spectral and spatial features vary with the incident laser power. Most notably, we can find two critical incident powers (turnpoints) at which the PL features substantially change, one at around 8.0 $kW/cm^2$ and the other at around 20.0 $kW/cm^2$. The high-power turnpoint at 20.0 $kW/cm^2$ features a substantial shrinking of the luminescence area (see the images of 19.9 $kW/cm^2$ and 21.2 $kW/cm^2$ in Fig. 1a as well as blue area of Fig.1c), a sharp increase in the PL intensity (see the orange (18.4 $kW/cm^2$) and grey (21.7 $kW/cm^2$) curves in Fig. 1b as well as blue area of Fig.1d), and a change of the PL width and peak position to be power independent (blue area of Fig.1e). According to the previous studies,[11] these changes in PL can be correlated to the formation of EHL due to the gas-liquid phase transition of excitons. In contrast, the low-power turnpoint at 8.0 $kW/cm^2$ features an expansion of the luminescence area (see the images of 6.51 $kW/cm^2$ and 8.01 $kW/cm^2$ in Fig. 1a and yellow area of Fig.1c), a change of the PL intensity from power-dependent to power-independent (yellow area of Fig.1d), and a larger broadening of the PL spectral width with increase of the incident power (yellow area of Fig.1e). We have correlated the low-power turnpoint to EMT in the previous studies,[11] and now focus on providing more detailed studies to further support the correlation and, more importantly, to illustrate the fundamentals of the EMT in this work.

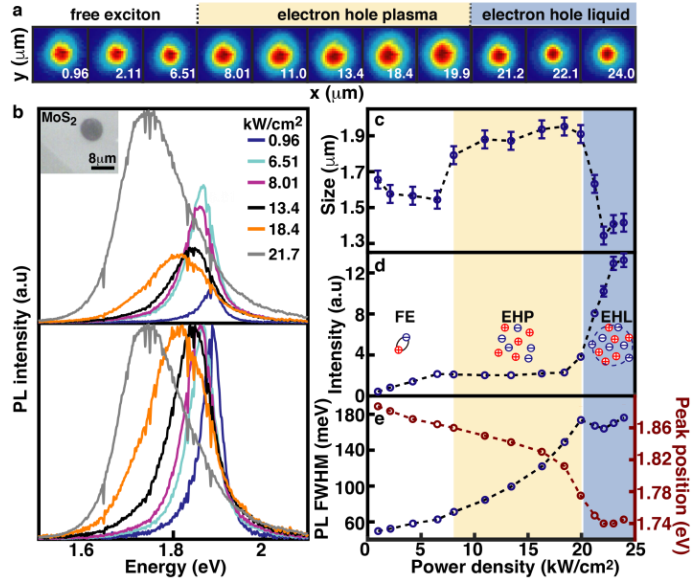

**Figure 1. Experimental evidence for exciton phase transition in suspended monolayer MoS$_2$.** **(a)** Representative spatial PL imaging collected from a suspended CVD-grown monolayer MoS$_2$ flake (in radius of 4.0 μm). The white number indicates the incident power density with a unit of kW/cm$^2$. **(b)** As-measured (top panel) and normalized (bottom panel) PL spectra collected from the luminescence center of monolayer MoS$_2$ under different incident power densities. Inset is an optical image for suspended monolayer MoS$_2$. **(c)** The size of the luminescence area (spatial full width at half magnitude), **(d)** the PL intensity, **(e)** the PL's peak position (red) and spectral full width half magnitude (FWHM, blue) as a function of the incident power density, which are extracted from the results in *(a)* and *(b)*. The areas with different colors in *(a)* and *(c-e)* indicate the different physical states of excitons, including free excitons (FEs), electron-hole plasma (EHP), and electron-hole liquid (EHP). Error bars are not plotted for the datapoints whose measurement error is smaller than the size of the dot.

**Evidence for exciton Mott transition**

To further support the correlation of the turnpoint at 8.0 kW/cm$^2$ to EMT, we examine and exclude out other possible factors being the major reason for the change of PL. These include laser-induced defects, temperature increase, and the increase of trions or biexcitons. First, the photothermal effect of incident laser can give rise to temperature increase at the monolayer, which may induce oxidation of the materials and generate defects if the temperature increase is high enough, and the presence of defects can affect the PL. However, we carefully controlled the measurement condition to minimize defect generation in experiment. The PL measurement was performed in Ar protected environment (see Methods) and the laser power utilized was mild, no more than 0.3 mW at the power density of 8.0 kW/cm$^2$ (the radius of the focused laser is 1.1μm ), which induces a temperature increase of no more than 140°C at the monolayer (Fig. 2a). According to our previous studies on suspended monolayer MoS$_2$,[11,24,25] such experimental conditions cannot generate observable defects at the monolayer. This is indeed supported by similar Raman/PL features of the monolayer before and after the power-dependent PL measurement (Fig. S1). Additionally, the change of PL at 8.0 kW/cm$^2$ is reversible, as the PL resumes the original low-power features when the laser power is lowered back (Fig. S2). This further supports that the change of PL at 8.0 kW/cm$^2$ is not caused by laser-induced defects, because the effect of defects, if any, are irreversible and can persist even after the laser power is lowered according to our previous studies.[29]

Second, the temperature increase of the monolayer induced by the incident laser can cause the PL to redshift and broaden, but we can exclude out the temperature increase being the mechanism for the change of PL at 8.0 kW/cm$^2$ as well on basis of quantitative analysis. Based on our previous studies on the thermal conduction of suspended monolayer MoS$_2$,[30] we can estimate the temperature increase of the monolayer induced by the incident laser as shown in Fig. 2a. With the information of the temperature increase, we can further estimate temperature-induced redshift and broadening of the PL by multiplying the temperature increase with a temperature-dependent redshift (0.317 meV/K) and broadening coefficient (0.19 meV/K) that we measured and plot in Fig. 2b-c. We plot the temperature-induced redshift and broadening along with the experiment observation in Fig. 2d-e. The result indicates that the temperature increase indeed accounts for the broadening and redshift of the PL at < 8.0 kW/cm$^2$, but cannot explain the PL's broadening and redshift at > 8.0 kW/cm$^2$, indicating that the laser-induced temperature increase is not the driving force for the change of PL at 8.0 kW/cm$^2$.

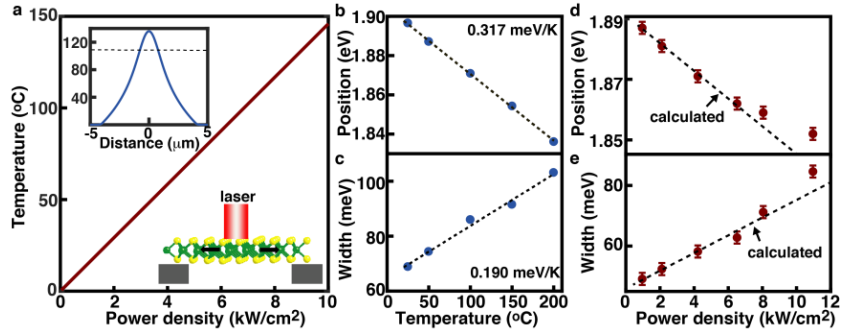

**Figure 2. Laser-induced temperature increase and its effect on PL.** (**a**) Calculated temperature increase at the monolayer as a function of incident lase power density. The lower inset is a schematic illustration for the illumination configuration. The upper inset is the calculated temperature profile of the monolayer at the incident power of 8.0 kW/cm$^2$, and the dashed line indicates the average temperature by weighting the spatial distribution of temperature as reported in Ref. 11. (**b**) Peak position and (**c**) spectral width of the emission of suspended monolaer MoS$_2$ measured at different temperatures. Measured (dots) and calculated (dashed line) (**d**) peak position and (**e**) spectra width as a function of incident power density.

Third, the PL of monolayer MoS$_2$ involves contribution from excitons and trions (charged excitons) even at low incident powers, and we can reasonably separate the contribution of the different excitonic species via fitting as shown in Fig. 3a. The fitting indicates that excitons play a dominant role in the PL. In principle, the increase of incident power may induce the formation of more trions or even bi-excitons, which can cause PL to obviously broaden and redshift when the emission intensity of trions or bi-excitons is comparable to or dominate over exciton emission.[31-33] However, we can exclude out the increase of trions or bi-excitons being the major reason for the PL change at 8.0 kW/cm$^2$. The PL intensity is pretty much independent of the incident power at > 8.0 kW/cm$^2$. This does not match the exponential power-dependent increase in intensity usually observed at trion or biexciton emission,[34] indicating that trions or biexcitons still do not play a major role in the PL at > 8.0 kW/cm$^2$. This conclusion is further supported by quantitative analysis of the PL spectra. Unlike the PL spectra at low incident powers, which features a sharp peak (Fig. 3a), the PL spectra at > 8.0 kW/cm$^2$ often involve a main peak accompanied by a weak but visible shoulder peak (Fig. 3b). Guided by the visible peaks, we can fit the PL spectra at the incident power 8.0-20.0 kW/cm$^2$ to a narrow peak at 1.86-1.87 eV (peak 1) and a broad peak at lower energies (peak 2) as shown in Fig. 3c. The narrow peak at 1.86-1.87 eV matches the exciton emission at the incident power right below 8.0 kW/cm$^2$ (see Fig. 2d) and can thus be ascribed to exciton emission. Interestingly, the broad emission shows distinct power dependence from the exciton emission (Fig. 3d-f). It shows substantial redshift (Fig. 3d), intensity increase (Fig. 3e), and broadening (Fig. 3f) with the incident power increasing, while the exciton emission shows decrease in intensity (Fig. 3e) and negligible change in peak position (Fig. 3d) and spectral width (Fig. 3f). The distinct power dependence from that of exciton emission indicates that the broad emission does not originate from trions or bi-excitons, because the emission of trions or bi-excitons is usually pinned with exciton emission,[31,34] including a constant energy difference in the peak positions and a similar trend (either power-dependent increase or decrease) in term of the dependence on incident power. In brief, while trions or bi-excitons are involved in our measurement, they do not play a major role in driving the change of PL at 8.0 kW/cm$^2$.

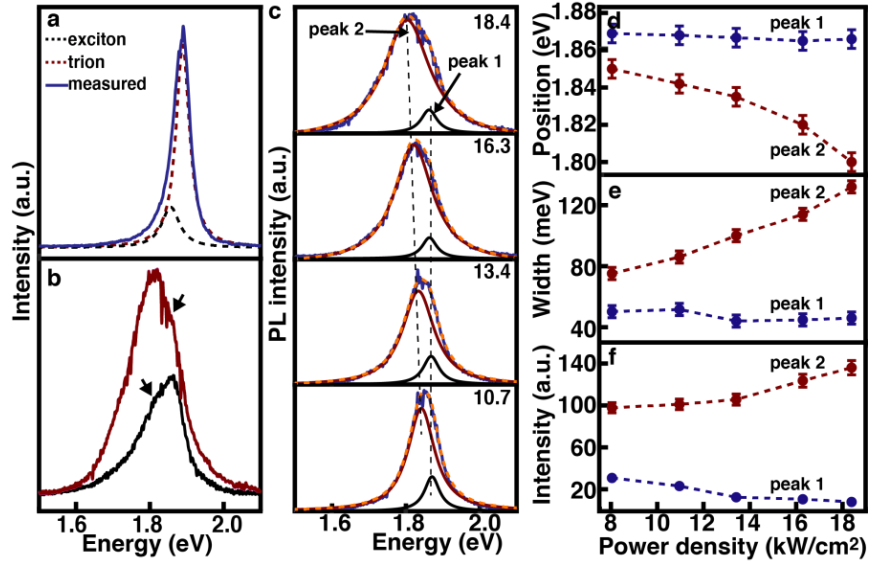

**Figure 3. Exclusion of the major role of trion/bi-exciton in the PL change at 8.0 kW/cm$^2$. (a)** measured (blue solid curves) and fitted (dashed line) PL collected from monolayer MoS$_2$ at the incident power 1.0 kW/cm$^2$. The fitting involves both contribution of excitons (red dashed curve) and trions (blue dashed curve). **(b)** Representative PL spectra collected from the monolayer at the incident power beyond 8.0 kW/cm$^2$. The arrows point towards shoulder peaks in the PL spectra. **(c)** PL spectra collected with different incident powers beyond 8.0 kW/cm$^2$ (as indicated by the given number with a unit of kW/cm$^2$) and corresponding fitting. The spectra are fitted as combination of one narrow peak at 1.86-1.87 eV (peak 1) and another broader one at lower energies (peak 2). The dashed lines illustrate the shift of the positions of peak 1 and peak 2. **(d)** Intensity, **(e)** peak position, and **(f)** spectral width of the fitted peaks (blue, peak 1; red, peak 2) as a function of the incident power.

Instead, the correlation of the PL change at 8.0 kW/cm$^2$ to exciton Mott transition (EMT) can reasonably account for all the experimental observation. Intuitively, the density of photogenerated charge carriers increases with the incident power increasing, and this enables increasingly screening for the pair-wise Coulomb attraction of excitons, which can eventually lead to the ionization of excitons into a plasma of unbound electrons and holes, *i.e.* electron-hole plasma (EHP). The correlation to EMT can account for the broad emission and its power dependence shown in Fig. 3d-f, as EHP is known featuring a broad emission at energies lower than exciton emission that substantially redshifts and broadens with increase of charge density. [8,11,15,17,35] The correlation may also account for the anomalous power-dependence observed at the exciton emission (Fig. 3d-f). The unbound charge carriers in EHP can screen the pair-wise Coulomb attraction of co-existing excitons, and this may lead to an increasing reduction in exciton binding energy and exciton emission efficiency with increase of the incident power, which gives rise to the observed decrease in exciton emission (Fig. 3f). The observed negligible redshift and broadening of the exciton emission with increase of the incident power (Fig. 3e), which is in stark contrast to the substantial thermally-induced redshift and broadening at lower incident powers (see Fig. 2d), can also be ascribed to the presence of EHP. The negligible redshift indicates that the bandgap renormalization and binding energy reduction induced by the Coulomb screening of the unbound charge carriers in EHP offset the photothermal effect on the peak position of exciton emission. The unbound charge carriers of EHP may also screen electron-phonon coupling, [36,37] which is a major mechanism for the thermally induced broadening in emission, and this can make the spectral width of exciton emission less sensitive to temperature increase and leads to the negligible broadening.

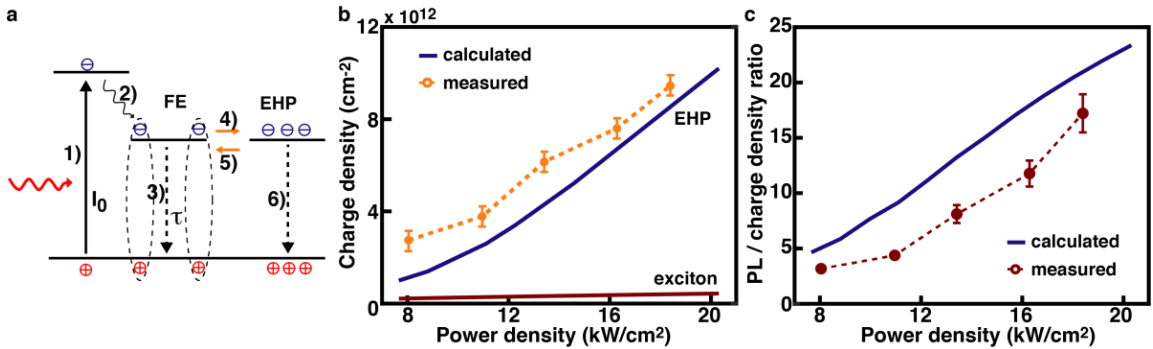

**Figure 4. Quantification of charge densities in co-existing EHP and exciton phases. (a)** Schematic illustration for the charge dynamics involved in exciton Mott transition. The process involves 1) photogeneration of charge carriers, 2) relaxing to band edges, 3) exciton recombination, 4) exciton ionization into EHP, 5) de-ionization of EHP to excitons, and 6) EHP recombination. **(b)** Measured (orange dots) and calculated charge densities in EHP (blue line) and excitons (red line) as a function of incident power. The measured charge densities are obtained on basis of the measured spectral width of EHP emission as shown in Fig.3e. **(c)** The calculated ratio of charge densities in the EHP and co-existing excitons (blue line) and the measured intensity ratio of EHP and exciton emission (red dots) as a function of incident power. The measurement result is obtained from the data shown in Fig. 3f.

**Quantification of the charge carrier distribution in EHP and excitons**

EHP and excitons co-exist upon the onset of EMT, and it is important to understand how the photo-generated charge carriers distribute among the two phases. But this information has remained elusive to date due to the difficulty in deterministically distinguishing the signals from different excitonic species in previous studies.[16,22] We explore to quantify the distribution of charge carriers among EHP and co-existing excitons by taking advantage of the reasonable separation of EHP and exciton emissions as shown in Fig.3d-f .

To do that, we examine the charge dynamic involved in EMT from intuitive perspective as illustrated in Fig. 4a. Excitons are generated upon photoexcitation (step 1), followed by a fast thermalization process (step 2) relaxing to the band edges. After that, the excitons decay through linear and nonlinear (exciton-exciton annihilation, EEA) recombination (step 3) as well as ionization into EHP (step 4). The resulting EHP decays via a de-ionization process back to excitons (step 5) and recombination (step 6). This dynamic process is complicated, but can be reasonably simplified on basis of a long-established fact that exciton recombination is much faster than exciton ionization as observed by us [8,24] as well as references[22,38-40]. We previously demonstrated that the lifetime of excitons in suspended monolayer $MoS_2$ is around 1 ns[24] and the ionization of excitons and the de-ionization of EHP are at the time scale of 100-300 ns.[8] Exciton ionization in other semiconductors such Si and quantum wells have also been demonstrated taking time as long as 150 ns.[22,38-40] As the recombination of excitons is much faster than the ionization/de-ionization process, it is reasonable to consider a two-stage model decoupling the dynamics of excitons and EHP in time. Upon photoexcitation, a steady state of excitons is reached first, then the excitons undergo a slow ionization process to build up charge carriers in EHP until a quasi-equilibrium between the excitons and EHP as $X \leftrightarrow e^- + h^+$ is reached, and the steady-state exciton density maintains unchanged during the ionization We can evaluate the steady-state exciton density without considering the ionization process as what we previously did for the scenario of low incident powers [41]

$$-\frac{n_X}{\tau_X} - kn_X^2 + D\nabla^2 n_X + q\nabla\cdot(\mu n_X \nabla E) + \alpha I_0 \exp\left(-\frac{2r^2}{w^2}\right) = 0 \quad \ldots(1)$$

The first two items represent the linear and nonlinear exciton recombination with $\tau_X$ being the linear recombination time and $k$ the EEA rate constant. The third term is exciton diffusion governed by exciton concentration gradient $\nabla n_X$ and diffusion coefficient $D$. The fourth term results from exciton funneling under the bandgap gradient induced by the photothermal effect of the incident laser, in which $\mu$ is exciton mobility, $q$ is the charge carried by single electrons, and $\nabla E$ is the photothermally induced bandgap gradient. $\alpha$ is the absorption efficiency of the monolayer for the incident laser, $w$ is the radius of the incident laser, and $I_0$ is the incident power density.

We estimate the steady-state exciton density as a function of incident power with eq. (1) using the parameters we measured in experiment ($\alpha = 0.065$,[42] $k = 0.04$ cm$^2$/s,[24,43] $\tau = 1.0$ ns,[24] $D = 22.5$ cm$^2$/s,, and $\mu$ depends on local temperature and the size of incident laser as reported previously[41]). The calculation result is plotted in Fig. 4b (red line). With the information of exciton density, we can estimate the charge density of the co-existing EHP with Saha equation, which has been extensively used to describe plasma ionization under thermal equilibrium.[44] We derive the Saha equation governing the ionization of excitons X ↔ e$^-$ + h$^+$ in 2D materials as

$$\frac{n_X}{n_P^2} = \frac{h^2}{2\pi k_B T}\frac{m_X}{m_e m_h}\frac{g_X}{g_e g_h}\exp\left(\frac{F_X - F_e - F_h}{k_B T}\right) \quad \ldots\ldots(2)$$

$m_i$, $g_i$, and $F_i$ represent the effective mass, valley degeneracy, and free energy of excitons ($i = X$), free electrons ($i = e$), and free holes ($i = h$), respectively. $h$ is the Planck's constant, and $k_B$ is the Boltzmann constant. $T$ is the local temperature of the monolayer. Most of the parameters in eq (2) are available in references ($m_X = 0.26\ m_0$, $m_e = 0.47\ m_0$, and $m_h = 0.60\ m_0$,[45] $m_0$ is the mass of static electrons; $g_X = g_e = g_h = 2$; the free energy of excitons $F_X$ is equal to the exciton binding energy $E_b$ ($E_b = 0.42$ eV[42]) but has an opposite sign. We can find out the local temperature $T$ of the monolayer following the result of our previous studies as illustrated at Fig. 2a.[30] We can also calculate the free energy of charge carriers with a well-established model for the free energy of electron-hole systems (see S1 for details of the model).[46] Based on that, we can find out the charge density $n_p$ of EHP with eq. (2) and plot the result in Fig. 4b (blue curve). The results indicate that the charge density in EHP rapidly increases by 10 folds from $1.0 \times 10^{12}$ cm$^{-2}$ to $1.0 \times 10^{13}$ cm$^{-2}$ in the incident power range of 8.0 - 20.0 kW/cm$^2$, while the charge density in the co-existing excitons only increases by 2 times from ~ $2.0 \times 10^{11}$ cm$^{-2}$ to ~ $4.0 \times 10^{11}$ cm$^{-2}$.

The quantitative estimate for charge carrier density in the co-existing exciton and EHP phases is reasonable, as it is consistent with well-accepted common sense, experimental observation, and theoretical prediction. First, the density of unbound charges carriers in EHP increases by 10 folds from $1.0 \times 10^{12}$ cm$^{-2}$ at the onset of EMT (~8.0 kW/cm$^2$) to $1.0 \times 10^{13}$ cm$^{-2}$ at the gas-liquid phase transition (~ 20.0 kW/cm$^2$). This is consistent with the common sense that the critical charge density for the onset of EMT is usually one order of magnitude lower than that for the gas-liquid phase transition.[16,22] Second, the estimated charge density in EHP is reasonably consistent with the charge density derived from experimental measurement (orange dots in Fig. 4b). As discussed later in Fig.7, the charge density of EHP can be derived from the spectral width of EHP emission (see Fig. 3e). Additionally, the ratio of the estimated charge densities in EHP and excitons reasonably matches the ratio of the measured PL intensities of EHP and excitons (see Fig.3f) as shown in Fig. 4c. Third, the estimated charge densities are consistent with theoretical prediction for the onset of EMT and exciton gas-liquid phase transition. Theoretical models predict that EMT starts when the Debye-Hückel screening length $L_s$ is comparable to the Bohr radius $a$.[22,47] The Debye-Hückel screening length of the unbound electrons and holes of EHP can be written as $L_s = 2\varepsilon k_B T/e^2(2n_p)$.[48] $\varepsilon$ is the dielectric constant of the monolayer ($\varepsilon = 2.85$ as derived from the Bohr radius $a$ (0.60 nm) and binding energy $E_b$ (0.42 eV) [42]). By taking into account the photothermal effect of the incident laser, we estimate the local average temperature $T = 414$ K at the incident power of 8.0 kW/cm$^2$ (as indicated by the dashed line in Fig. 3a). Therefore, the critical charge density to enable EMT in monolayer MoS$_2$ is predicted to be $0.94 \times 10^{12}$ cm$^{-2}$, which is consistent with the estimated charge density $1.0 \times 10^{12}$ cm$^{-2}$ for the EHP at 8.0 kW/cm$^2$. Additionally, the gas-liquid phase transition is predicted to occur when the chemical potentials of EHP and EHL are equal. Theoretical analysis further predicts that the critical charge density of EHP to enable the gas-liquid phase is indeed $1.0 \times 10^{13}$ cm$^{-2}$ [49], nicely matching the estimated charge density in EHP at 20.0 kW/cm$^2$ at which we observe the gas-liquid phase transition (see Fig. 1).

We can get more insight into the charge carrier distribution among excitons and EHP by further examining the two-stage model from intuitive perspective. The two-stage model is developed on basis of experimental observation that exciton ionization is two orders of magnitude slower than exciton recombination. It considers that a steady state of excitons is formed first, followed by ionization of the excitons to build up unbound charges in EHP until a quasi-equilibrium between the excitons and EHP is reached. It also assumes that the steady-state exciton density does not change during the ionization process. The assumption is reasonable given the consistence of the calculation results with common sense, experimental observation, and theoretical prediction as discussed above. Intuitively, we can consider that the loss of excitons due to the ionization may be well compensated by the injection of new excitons via photoexcitation.

The correlation of excitons with EHP is similar to the correlation of excitons with trions, the former governed by the thermodynamics of exciton ionization $X \leftrightarrow e^- + h^+$ and the latter by the thermodynamics of exciton-charge combination $X + e\,(h) \leftrightarrow X^{-(+)}$. However, unlike EHP, the formation of trions can substantially affect the density of excitons, because the dynamics of trion formation cannot be decoupled from the dynamics of excitons as the exciton-charge combination rate is comparable to the rate of exciton recombination.[50]

**Clarification for the controversy/ambiguity of EMT: continuousness and criteria**

Our experimental results provide clarification for two long-standing controversies about EMT, whether it is continuous or abrupt and the critical charge density for the onset of EMT. The results in Fig. 1 and Fig. 3 show that the EHP emission continuously evolves with the incident power increasing up to around 20.0 kW/cm$^2$ at which the exciton gas-liquid phase transition starts. This indicates that the EMT in monolayer MoS$_2$ is a continuous process. As schematically illustrated in Fig. 4a, the EMT starts at the incident power of around 8.0 kW/cm$^2$, continuously evolves with further increase of the incident power, and terminates by transiting into the gas-liquid phase transition at around 20.0 kW/cm$^2$. In contrast to the continuous nature of the EMT, the gas-liquid phase transition in monolayer MoS$_2$ is demonstrated to be a first-order, abrupt process.[11] As illustrated in Fig.1a., the major change of PL associated with the gas-liquid phase transition finishes within a range of 2.0 kW/cm$^2$, and this range is actually dictated by the spatial/spectral resolution of the characterization technique rather than the nature of the gas-liquid phase transition.

The continuous nature of EMT generally holds for other semiconductors as reported in many references. Whereas some previous studies did argue that EMT is a first-order, abrupt transition,[51] this might due to mistakenly considering exciton gas-liquid phase transition to be EMT. We find that the changes in PL ascribed to EMT in those studies[51] are actually very similar to the well-recognized spectral features associated with exciton gas-liquid phase transition, for instance, the PL peak position and spectral width showing no change with increase of incident power.[46] The mistaken consideration of exciton gas-liquid phase transtion to be EMT is not uncommon at all. For instance, there is a widely contracted mistake in references that uses the physical touch of neighboring excitons as a criteria for the onset of EMT, such as $n^2a = 1$ (or 1/4) for 2D materials and $n^3a = 1$ (or 1/8 or 0.26) for 3D materials or similar equations,[52] but the physical touch of neighboring excitons is actually a well-established criteria for exciton gas-liquid phase transition as reported in other references.[46] For instance, we demonstrated that the excitons in monolayer $MoS_2$ reasonably touch each other upon the gas-liquid phase transition, as the charge density and exciton Bohr radius bear a relationship of $n^2a = (1/2.6)^2$.[11]

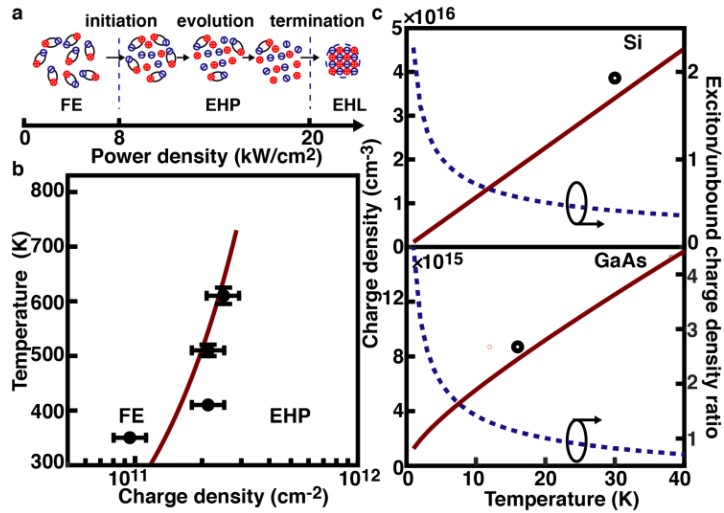

**Figure 5. Continuousness and criteria of EMT.** (**a**) Schematic illustration for the continuousness of exciton Mott transition, starting at 8.0 kW/cm$^2$, evolving with increase in incident power, and terminating by merging into the gas-liquid phase transition at 20.0 kW/cm$^2$. (**b**) Calculated (solid curve) and measured (black dots) critical exciton density for the onset of EMT in monolayer MoS$_2$ at different temperatures of the monolayer. The temperature is a sum of the environment temperature plus laser-induced temperature increase. (**c**) Calculated critical density (red lines) of unbound charges for the onset of EMT in Si (top) and GaAs (bottom) at different temperatures. The dashed curves are the calculated ratio of the densities of excitons and unbound charges ($n_X/n_p$) at the onset of EMT. The result of Si is for unbound charges $n_p$, but that of GaAs is for total charges $n_X + n_p$, which is for convenience to compare with the experimental results in references[53,54].

Our experimental results also clarify the long-standing ambiguity in the criteria for the onset of EMT. EMT has been well recognized to start when the Debye-Hückel screening length $L_s$ is comparable to the Bohr radius $a$,[16,47] but it has never been specified which species should be counted in the evaluation of screening length, either excitons or unbound charges. Previous studies often do not explicitly specify if the charges are unbound charges or excitons or both during the evaluation of critical charge density. This ambiguity is one major source for the inconsistence in references regarding the critical charge density for the onset of EMT. Our result shows that the density of unbound charges at the onset of EMT derived from experimental measurement is consistent with the critical charge density predicted by the Debye-Hückel screening length. This indicates that the screening effect of unbound charges play a dominant role in driving the ionization of excitons, which is intuitively understandable because unbound charges are known able to provide stronger Coulomb screening than excitons and the equation of Debye-Hückel screening length is developed specifically for the screening of free charges.

On basis of the clarification, we develop an approach to find out the proper experimental conditions to enable EMT. We first estimate the critical density of unbound charges for the onset of EMT with $n_p \approx \varepsilon k_B T/e^2 a$ according to the criteria of Debye-Hückel screening. Since the control of exciton density is more straightforward in experiment, we find out the density of excitons under quasi-equilibrium with the critical density of unbound charges using Saha equation as shown in eq.(2). From the perspective of thermodynamic equilibrium, the free energies of excitons and unbound charges are comparable as $F_X \approx F_e + F_h$ upon the onset of EMT. Without losing generality, we take into account thermal energy, which may facilitate exciton ionization, and set the energy difference $F_X - F_e - F_h$ at the onset of EMT to be comparable to the thermal energy as $F_X - F_e - F_h = k_B T$ (certain variation of this value won't qualitatively change our conclusion). Then we can have the exciton density under quasi-equilibrium with the critical density of unbound charges, which is referred as critical exciton density

$$n_X = \left(\frac{\varepsilon h}{2\pi e^2 a}\right)^2 \frac{2\pi k_B m_X g_X}{m_e m_h g_e g_h} T \exp(-1) \quad \ldots\ldots (3)$$

Fig. 5b shows the calculated critical exciton density for the onset of EMT in monolayer MoS$_2$ as a function of temperature. To validate the calculation result, we performed power-dependent PL measurement at suspended monolayer MoS$_2$ under different environment temperatures, and estimated the exciton density with eq. (1) as well as the laser-induced temperature increase of the monolayer at the onset of EMT as what we did in Fig. 2 and Fig. 4. We plot the exciton density and local temperature of the monolayer derived from experimental measurement in Fig.5b. The experimental results show reasonable consistent with the theoretical calculation.

The approach can also be applied to estimating the critical charge densities for the onset of EMT in conventional three-dimensional (3D) semiconductors. Using the Debye-Hückel screening and Saha equation for 3D materials, we can have the critical densities of unbound charges and excitons as

$$n_p \approx \varepsilon k_B T / 2e^2 a^2 \quad \ldots\ldots (4)$$

$$n_X = n_p^2 \frac{h^3}{(2\pi k_B T)^{1.5}} \left(\frac{m_X}{m_e m_h}\right)^{1.5} \frac{g_X}{g_e g_h} \exp(-1) \quad \ldots\ldots (5)$$

We performed the evaluation for Si and GaAs with parameters available in references (for Si, $a = 4.4$ nm, $\varepsilon = 11.4$, $m_e = 0.49\, m_o$, $m_h = 0.19\, m_o$; for GaAs, $a = 11.6$ nm, $\varepsilon = 12.9$, $m_e = 0.067\, m_o$, $m_h = 0.082\, m_o$; $m_X = (1/m_e + 1/m_h)^{-1}$)[55]. Fig. 5c shows calculated critical charge density (solid lines) as well as the ratio of the critical densities of excitons and unbound charges ($n_X/n_p$, dashed curves) as a function of temperature. The calculations results are reasonable consistent with the critical charge density measured in experiments (black circles in Fig. 5c).[53,54] For instance, for Si at 30 K $n_p$ = 3.9 ×10$^{16}$ cm$^{-3}$ (experimental), $n_p$ = 3.4 ×10$^{16}$ cm$^{-3}$ (calculation); for GaAs at 16 K, $n_p + n_X$ = 8.7 ×10$^{15}$ cm$^{-3}$ (experimental, 2.0 ×10$^{10}$ cm$^{-2}$ quantum wells divided by the thickness of 23 nm), $n_p + n_X$ = 7.8 ×10$^{15}$ cm$^{-3}$ (calculation).

**Physical state of the charge carriers in EHP: individually unbound or complex?**

The unbound charge carriers in EHP are usually believed existing individually without much inter-charge interactions, similar to free charges. However, our experimental results indicate that the charge carriers in EHP exist in format of electron-hole complex in nanometer-scale size instead, which we refer as *EHP bubble*. This is evidenced by different spatial distributions in the PL intensity and spectral features (peak position and spectral width) of EHP. Without losing generality, we use the result collected at 18.4 kW/cm$^2$ as an example to illustrate this notion. Figure 6a shows two-dimensional (2D) spatial-spectral images collected from monolayer MoS$_2$. To illustrate the spatial distribution of the spectral features, we normalize the PL spectrum at every location with respect to the corresponding peak intensity and plot the normalized spatial-spectral image at the right panel of Fig. 6a. The PL emission involves contribution from both EHP and excitons as discussed in Fig.3. We can separate the emission from EHP and excitons via fitting that follows the guidance of visible peaks, as illustrated in Fig. 6b. The fitting results, including PL intensity, spectral width, and peak position of the EHP and exciton emissions are plotted as a function of the distance away from the luminescence center in Fig. 6c. Significantly, the PL intensity of EHP substantially decreases by 5 folds with increase of the distance (top panel of Fig. 6c), but the peak position and spectral width only shows very mild variation of < 20 meV (middle and bottom panels of Fig. 6c). To intuitively illustrate the mild spatial variation of the spectral features of EHP, we plot normalized PL spectra at different locations in the top panel of Fig.6d.

We find the low energy tails of all the PL spectra, which is mainly contributed by EHP emission, nicely overlap as indicated by the horizonal dashed line. As a comparison, we re-plot all the PL spectra collected from the luminescence center with different incident powers (originally plot in Fig.3c) at the bottom panel of Fig.6d. The low energy tails of these spectra are well separated due to the blueshift and narrowing of EHP emission in these spectra as illustrated in Fig. 3d-f.

The different spatial distributions of the PL intensity and spectral features of EHP is counter intuitive, as they point out contradictory spatial distribution of the charge density. The 5-fold decrease in intensity suggests a substantial decrease in the charge density with the distance increasing, but the minor change in peak position and spectral width suggests only a mild variation in the charge density over the same distance. This conflict can be conciliated by considering that the charge carriers in EHP exists in format of small clusters (referred as *EHP bubble*) rather than individual free charges, as illustrated in Fig. 6e. The PL intensity is determined by the total number of charge carriers, which is equal to the number of the bubbles multiplying the charge density in each bubble, while the spectral width and peak position are only dictated by the charge density in each bubble. The experimental observation indicates that the total number of charge carriers in EHP decreases by 5 folds across the luminescence area, but the charge density in each bubble does not change much. Additionally, the similar charge density in all the EHP bubbles across the luminescence area indicates that the EHP bubbles are generated at the luminescence center and then diffuse over the luminescence area.

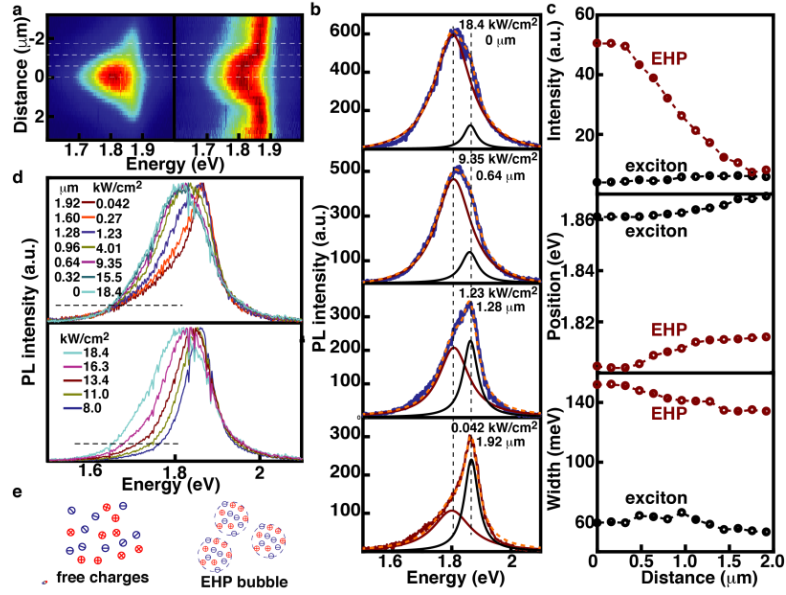

**Figure 6. Evidence for EHP bubbles. (a)** As-measured (left) and normalized (right) two-dimensional spectral-spatial PL image of monolayer MoS$_2$ collected at the incident power of 18.4 kW/cm$^2$. In the normalized result, the PL spectrum at each position is normalized with respect to the local maximum intensity. Horizontal dashed lines indicate the location of the four representative spectra given in (*b*). **(b)** Measured and fitted PL spectra collected at different locations as indicated by the dashed horizontal lines in (*a*). The dashed lines connect the peak positions of the EHP and exciton emission. **(c)** Fitted intensity, position, and spectral width of the exciton emission (black) and EHP emission (red). **(d)** Comparison of the PL spectra collect from different locations at 18.4 kW/cm$^2$ (top) and the PL spectra from the luminescence center at different incident powers (bottom). The legend in the top panel indicates the distance of each spectra away from the center and corresponding local incident power density. The dashed horizonal lines indicate the low energy tails of the spectra. **(e)** Schematic illustration for individual unbound charges and EHP bubbles.

We can roughly estimate the size of EHP bubbles on basis of diffusion coefficient as what reported previously.[56] Again, we use the result collected at 18.4 kW/cm$^2$ as an example to illustrate this notion. From Fig. 6c, we can roughly estimate the diffusion length $L_{EHP}$ of the EHP bubbles to be around 1.2 µm. We previously demonstrated that the lifetime $\tau_{EHP}$ of EHP under comparable conditions is around 100 ns.[8] Therefore, the diffusion coefficient $D_{EHP}$ of the EHP bubble can be estimated to be around 0.15 cm$^2$/s on basis of $D_{EHP} = L_{EHP}^2/\tau_{EHP}$. On the other hand, we find the diffusion coefficient of excitons to be 22.5 cm$^2$/s with the presence of EHP.[36] Because the diffusion coefficient is inversely proportional to the mass $M$, we can estimate the number of electron-hole pairs in the EHP bubble to be around 150, which corresponds to a size of around 50 nm at the charge density of $0.88 \times 10^{13}$ cm$^{-2}$ at 18.4 kW/cm$^2$ (see Fig. 4b). This likely over-estimates the charge number and size of the bubble due to the simplicity of the estimate approach. Nevertheless it provides useful insight into the physical state of EHP charges. We also expect the size and number of charges in EHP bubbles to vary with incident power, which would need more studies for better understanding.

**Correlation of the emission spectral feature and charge density of EHP**

One barrier that prevents the quantitative studies of EMT lies in the lack of quantitative correlation between the PL spectral features and charge density of EHP, although it has been qualitatively understood that the PL spectral features of EHP is associated with charge density. The lack of the correlation makes it difficult to quantitatively evaluate the charge density of EHP, which is critical for the studies of EHP. This is in stark contrast with EHL, whose emission lineshape is well recognized to be the convolution of the occupied electron and hole density of states at the conduction and valence bands, respectively.[16,22] Previous studies also tried to assume the emission profile of EHP to be the convolution of the occupied electron and hole density of states as well, but the result is not consistent with experimental observation.[16,22]

We find that EHP emission can be well fit with a Lorentzian oscillator $(\Gamma/2)^2/[(E-E_0)^2+(\Gamma/2)^2]$ as illustrated in Fig. 3c and Fig. 6b, where $E_0$ is the peak position and $\Gamma/2$ is a broadening term that dictates the FWHM (full width at half magnitude) $\Gamma$ of the emission peak. We also find that the broadening of the EHP emission $(\Gamma/2)_{EHP}$ is reasonably equal to a sum of the broadening of the exciton emission $(\Gamma/2)_{ex}$ and the Fermi energies of the EHP as $(\Gamma/2)_{EHP} = (\Gamma/2)_{ex} + E_{F,e} + E_{F,h}$. Fig.

7b plots the measured difference in spectral broadening between the EHP emission and the exciton emission $(\Gamma/2)_{EHP} - (\Gamma/2)_{ex}$ (extracted from Fig. 3f). Also plotted is the Fermi energy of the EHP as $E_{F,e} + E_{F,h} = n\pi\hbar^2/2m_X$ calculated on basis of the charge density of EHP given in Fig. 4b. The calculated Fermi energy shows reasonable consistence with the measured difference in spectral broadening. Additionally, we find that the peak position $E_0$ of EHP is determined by the electronic bandgap $E_{eBG}$ of monolayer $MoS_2$ and EHP's free energy $F_{EHP}$, $E_0 = E_{eBG} + F_{EHP}$. As schematically illustrated in Fig. 7a, this is similar to exciton emission, whose peak position is dictated by the electronic bandgap $E_{eBG}$ (2.31 eV[42]) and exciton binding energy $E_b$, the latter of which is equal to excitons' free energy $F_X$ but has an opposite sign $E_b = -F_X$. We calculate the free energy of EHP using the well-established model for the free energy of electron-hole systems,[46] which involves Fermi energy, bandgap renormalization, and entropy associated with the charges (see S1 for details of the model). Fig.7c shows the measured and calculated peak position $E_0$ of the EHP emission. The calculation reasonably reproduces the measured peak position and its trend of redshifting with the incident power increasing. The non-trivial deviation at relatively low incident powers is rooted in the intrinsic imperfection of the model, which involves calculation of entropy on basis of statistic thermodynamics that works the best for high charge densities (see S1 for details of the model).

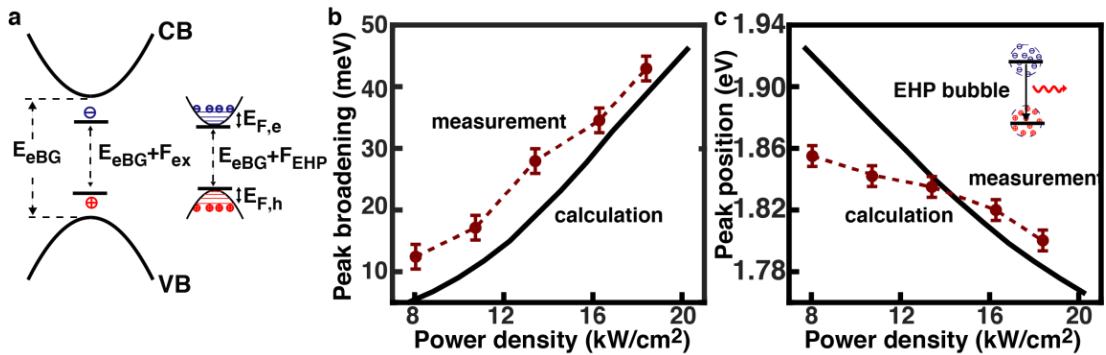

**Figure 7. Correlation of the PL spectral features and charge density of EHP.** (a) Schematic illustration for the free energy and Fermi energy of EHP as well as its correlation with the emission peak position. The correlation between the peak position and the free energy of excitons (which is equal to the exciton binding energy) is also illustrated. **(b)** The calculated Fermi energy of EHP (black) and the measured difference in the broadening (half of the spectral width) of EHP emission and exciton emission (red) as a function of incident power. **(c)** The calculated (black) and the measured (red) peak position of EHP emission as a function of incident power. The calculated peak position is equal to the sum of the electronic bandgap (2.31 eV) and EHP's free energy ($E_{eBG}$ + $F_{EHP}$, the free energy is negative). Inset, schematic illustration for the emission of EHP bubble.

The quantitative correlation of the spectral features and charge density of EHP provides a useful way to evaluate the charge density of EHP from PL spectral features. It also provides further insight into the physical state of the charge carriers in EHP. The reasonable fitting of EHP emission with a Lorentzian oscillator indicates that the EHP emits like an electron-hole complex as illustrated in Fig. 7c inset. The transition occurs at the energy level defined by the free energy of the EHP, and the large number of charges in EHP, which occupy multiple energy levels, simply increases the uncertainty in the transition energy to broaden the emission spectra. The observed emission lineshape is different from that expected from the emission of individually free charges, like the emission of EHL, which would show an asymmetric and narrower lineshape matching the convolution of the occupied electron and hole density of states.[46] This further support that the charge carriers in EHP do not exist as individual free charges, but as electron-hole complex instead. Given that the charge carriers in EHP emit like electron-hole complex rather than a plasma of completely free charges, we believe that *exciton plasma (EP)* is a better term to describe the excitonic phase resulting from EMT than electron-hole plasma (EHP).

**Conclusion**

In conclusion, we have elucidated new numerous insights into the fundamentals of EMT by taking advantage of monolayer $MoS_2$. These include clarification for the controversy on the continuous nature and criteria of EMT, quantification for the charge carrier distribution among the co-existing exciton and EHP phase, discovery of the physical state of charge carriers in EHP as nanoscale electron-hole complex (referred as *EHP bubbles*), and correlation between the emission profile and charge density of EHP. The result lays down a foundation to use EMT in quantum science and to development devices with broad implication in lasers, LEDs, imaging sensors, and optical interconnects.

**Methods**

*Synthesis and transfer of monolayer MoS$_2$:* The monolayer was grown on SiO$_2$/Si substrates using a chemical vapor deposition process,[26] and then was transferred onto the quartz substrates with pre-patterned holes in size of 4μm using a surface-energy-assisted transfer approach.[28] The holes were fabricated using standard photolithography and dry etching processes. In a typical transfer process, 9 g of polystyrene (PS, MW=280 kg/mol) was dissolved in 100 mL of toluene. The PS solution was spin-coated (3000 rpm for 60 s) on the as-grown monolayer. This was followed with 85° C baking for 1 hr. A water droplet was then dropped to assist the lift off of the PS-monolayer assembly. The polymer-monolayer assembly was then transferred to the quartz substrate. The sample was baked for 30 minuets at 150° C. The PS was removed by rinsing with toluene and followed by bake at 80° C for 30 minuets.

*PL measurement*: PL measurements were performed using a focused 532 nm laser in radius of 1.10μm. PL was collected by a long distance 50× objective and imaged through a long-pass filter with laser reflection light filtered out onto the entrance slit of a spectrometer. The spectrometer output was recorded with a PIXIS CCD camera. The PL imaging was recorded by moving the grating to the 0$^{th}$ order mode. In this collection mode, the spatial PL intensity distribution will be imaged onto the CCD camera at once. The PL spectrum was recorded by moving the grating to the 1$^{st}$ order diffraction mode. All the PL measurements were done with the samples placed in a chamber flown with Ar gas.

**Supplementary Materials:**

Figures S1-S3

S1. Calculation of Fermi energy and free energy of charge carriers

**Acknowledgments**

This work was supported by the National Science Foundation under a grant of EFMA 1741693.

**Author contributions**

Y. Y. and L.C. conceived the idea, Y. Y. performed the measurement, Y. Y. and L.C. designed the experiments, analyzed the data and wrote the manuscript. and analyzed the data. G. L. helped with the synthesis and transfer of the samples. All the authors were involved in reviewing the manuscript.

**Competing financial interests**

The authors declare no competing financial interests.